\documentclass[aps,prl, amsmath, amssymb, reprint, superscriptaddress]{revtex4-1}

\usepackage{graphicx}
\usepackage{dcolumn}
\usepackage{bm}
\usepackage{amssymb}
\usepackage{subfigure}
\usepackage{soul}
\usepackage{color}
\usepackage[normalem]{ulem}
\usepackage{todonotes}
\usepackage{color}

\begin{document}

\preprint{Submitted to Physical Review Letters}

\title{A frequency-preserving and time-invariant metamaterial-based nonlinear acoustic diode}

\author{A. S. Gliozzi}
\email{antonio.gliozzi@polito.it}
\affiliation{Department of Applied Science and Technology, Politecnico di Torino, Corso Duca degli Abruzzi 24, 10129 Torino, Italy}

\author{M. Miniaci}
\altaffiliation[Also at: ]{EMPA, Laboratory of Acoustics and Noise Control, \"Uberlandstrasse 129, 8600 D\"ubendorf, Switzerland}
\affiliation{D. Guggenheim School of Aerospace Engineering, Georgia Institute of Technology, North Avenue, GA 30332 Atlanta, USA}


\author{A. O. Krushynska}%
\affiliation{Department of Physics, University of Torino, Via Pietro Giuria 1, 10125 Torino, Italy}

\author{B. Morvan}
\affiliation{University of Le Havre, Laboratoire Ondes et Milieux Complexes, UMR CNRS 6294, 75 Rue Bellot, 76600 Le Havre, France}

\author{M. Scalerandi}%
\affiliation{Department of Applied Science and Technology, Politecnico di Torino, Corso Duca degli Abruzzi 24, 10129 Torino, Italy}

\author{N. M. Pugno}%
\altaffiliation[Also at: ]{School of Engineering and Materials Science, Queen Mary University of London, Mile End Road, London E1 4NS, United Kingdom;
Ket Lab, Edoardo Amaldi Foundation, Italian Space Agency, Via del Politecnico snc, 00133 Rome, Italy}
\affiliation{Laboratory of Bio-Inspired and Graphene Nanomechanics, Department of Civil, Environmental and Mechanical Engineering, Università di Trento, via Mesiano, 77, I-38123 Trento, Italy}

\author{F. Bosia}%
\email{federico.bosia@unito.it}
\affiliation{Department of Physics, University of Torino, Via Pietro Giuria 1, 10125 Torino, Italy}

\date{\today}

\begin{abstract}

We present the realization of an acoustic diode or rectifier, exploiting symmetry-breaking nonlinear effects like harmonic generation and wave mixing and the filtering capabilities of metamaterials. The essential difference and advantage compared with previous acoustic diode realizations is that the present is simultaneously a time invariant, frequency preserving and switchable device. This allows its application also as an \textit{on-off} or \textit{amplitude-tuning} switch. We evaluate its properties by means of a numerical study and demonstrate its feasibility in a preliminary experimental realization. This work may provide new opportunities for the practical realization of structural components with one-way wave propagation properties.

\end{abstract}


\maketitle

\section{Introduction}
In acoustics, the invariance of the wave equation with respect to time inversion, also known as reciprocity, has been exploited in many applications, e.g. for wave focusing in Time Reversal experiments \cite{Fink2000}. However, reciprocity is not necessarily desirable in all cases, especially when the goal is to isolate a source from its echos. Removal of unwanted reflections could indeed find numerous applications, such as acoustic one-way mirrors to prevent an ultrasound source from being disturbed by reflected waves \cite{Liang2009,Liang2010},  unidirectional sonic barriers to block environmental noise in a predefined direction \cite{Li2010}, control of acoustic energy transmission in medical applications using focused ultrasound \cite{Haar2006}, and energy harvesting \cite{Liu2016}. To achieve this, researchers in the field of acoustics and ultrasonics have drawn inspiration from electromagnetism, in the quest for a simple and efficient realization of an Acoustic Diode (AD) or rectifier. However, as illustrated by Maznev et al. \cite{Maznev2013}, linear elastic systems cannot be exploited to create ADs or isolators because they do not violate the reciprocity principle, so that the symmetry needs to be broken, for instance by periodically varying the elastic properties in space and/or time or by means of the introduction of nonlinearity coupled with some other mechanism (e.g. attenuation) \cite{Maznev2013}.

Examples of the first approach (periodically varying elastic properties in space and time) are provided by theoretical and numerical studies of one-dimensional system described by the discrete nonlinear Schr\"{o}dinger equation with spatially varying coefficients embedded in a linear lattice \cite{Lepri2011}, of continuous elastic systems with periodically-modulated elastic properties in space and time \cite{Trainiti2016}, or of a non-reciprocal active acoustic metamaterial \cite{Popa2014}.

Examples of the second approach (introduction of nonlinearity), instead, include 1-D design of a ``superlattice'' structure coupled with a nonlinear elastic medium \cite{Liang2009}, later realized experimentally using a contrast agent microbubble suspension to generate the nonlinearity \cite{Liang2010}, converting energy from the fundamental frequency to higher harmonics \cite{Maznev2013}.
Since then, several experimental realizations of ADs or rectifiers based on different mechanisms have been achieved. For instance, in \cite{Li2011}, unidirectional transmission was obtained through mode conversion, using a sonic crystal, rather than elastic nonlinearity. In \cite{Feng2014}, a mechanical energy switch and transistor are implemented by exploiting nonlinear dynamical effects of a granular crystal chain. To break the transmission symmetry, Ref. \cite{Fleury2014} proposed to use a subwavelength acoustic resonant ring cavity filled with a circulating fluid, splitting the degenerate azimuthal resonant modes, in analogy with the Zeeman effect in electromagnetism. In \cite{Sun2012}, a thin brass plate with single-sided periodical gratings immersed in water was shown to provide unidirectional transmission in a broad frequency range. Finally, in \cite{Zhu2016}, a passive multi-port structure with asymmetric sound transmission between neighbouring ports was presented. Comprehensive reviews of these and other approaches can be found in \cite{Maznev2013, Sklan2015}, in the latter with special reference to information processing in phononic computing, while the optimization of a rectifier efficiency in periodic mass--spring lattices is discussed in \cite{Ma2013}.

Many of these approaches are based on designing periodic structures, mainly phononic crystals and elastic metamaterials, which have attracted much attention for their wave manipulation capabilities, including negative refraction \cite{Morvan2010}, frequency Band Gap (BG)  formation \cite{Kushwaha1993,Martinez1995,Fraternali2017}, wave filtering or focusing \cite{Yang2004, Brun2010, Gliozzi2015, Miniaci2017}, scattering free propagation \cite{Miniaci2018} and acoustic cloaking \cite{Zhang2011}. Recent studies have shown how structural instabilities induced in ``static'' mechanical metamaterials can be exploited to achieve highly nonlinear dynamic response  that can be tailored to requirements \cite{Pasini2015,Bertoldi2017} and how weakly nonlinear monoatomic lattice chains can provide active control on elastic waves in phononic crystals \cite{Wang2016}. These or other approaches can be exploited to generate the type of nonlinearity required to violate spatial reciprocity  in elastic wave propagation \cite{Scalerandi2012}. On the other hand, phononic crystals and metamaterials are ideal candidates to efficiently realise large BGs \cite{Deymier2013,Krushynska2017} or to concentrate energy in to selected frequency ranges \cite{Carrara2013, Miniaci2017}.

In this paper, we propose the realization of an AD, based on the use of linear phononic crystals and elastic metamaterials,  embedded between elastic nonlinear regions. The novelty of the device is that it is simultaneously time--invariant (in the sense that its physical properties are not modified externally from the forward to the backward propagation direction \cite{Tsakmakidis2017}) and frequency preserving. Furthermore, besides its functionality as a diode, for other applications the device could be activated or deactivated at will, transforming it into a switch with the additional possibility to tune the amplitude of the output signal. These characteristics are in general not concurrently present in other AD designs that exploit nonlinearity to break the propagation symmetry and to transfer energy from the fundamental to the harmonics, with a frequency variation from input to output. The originality of our approach also resides in the exploitation of the combined effects of two different features of nonlinear elastic wave propagation, i.e. higher order harmonic generation and wave mixing, which allow to preserve the operating frequency taking place in two different zones separated by the periodic (filtering) structure. Wave mixing occurs when two longitudinal waves propagating through a nonlinear elastic zone interact and generate another longitudinal wave with a frequency given by the difference (and sum) of the frequencies of the two original waves. 

\section{Model and Method}

\begin{figure}
\centering
\begin{minipage}[]{1\linewidth}
\subfigure
{\includegraphics[trim=10mm 90mm 130mm 60mm, clip=true, width=.98\textwidth]{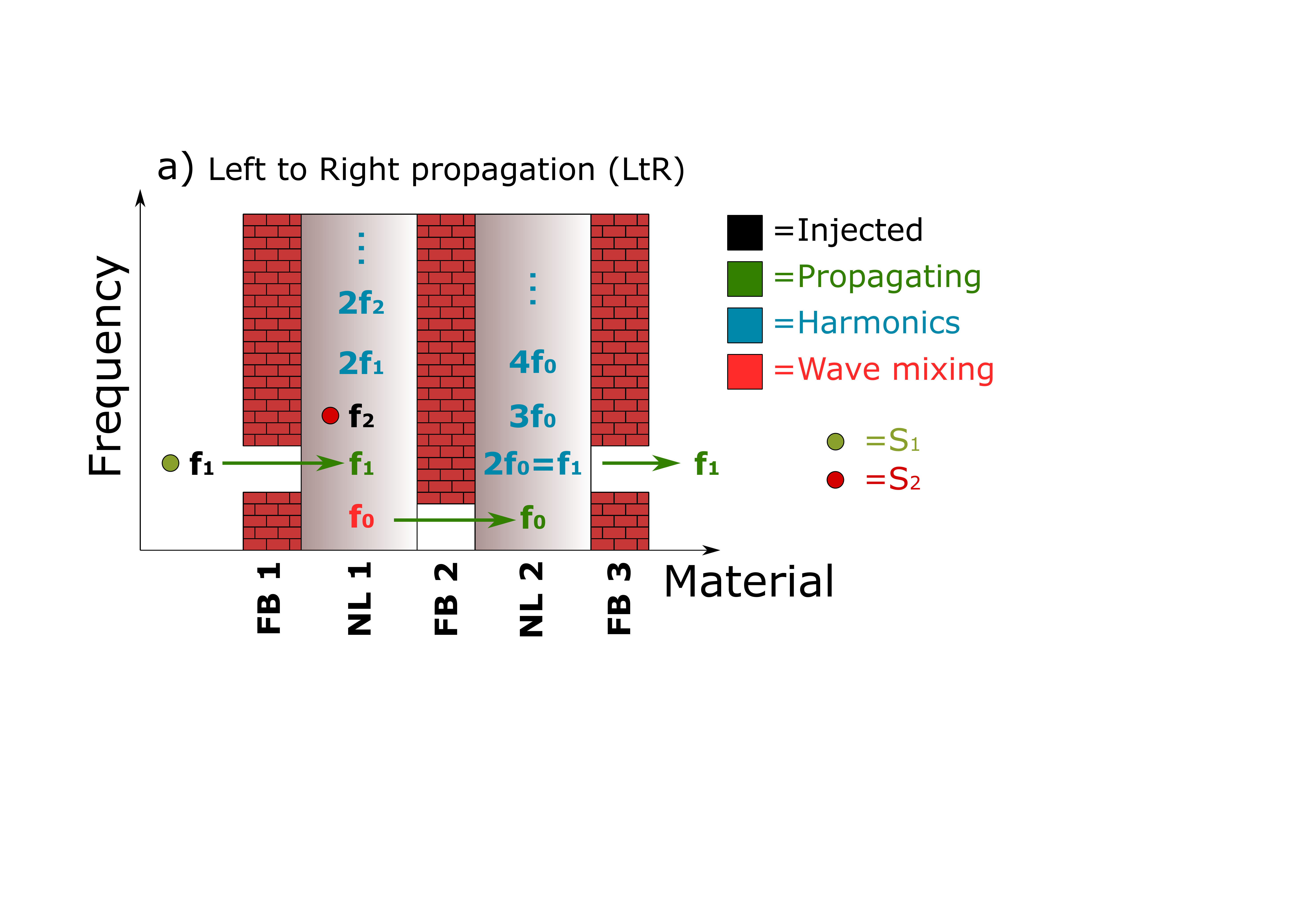}}
\subfigure
{\includegraphics[trim=10mm 90mm 130mm 60mm, clip=true, width=.98\textwidth]{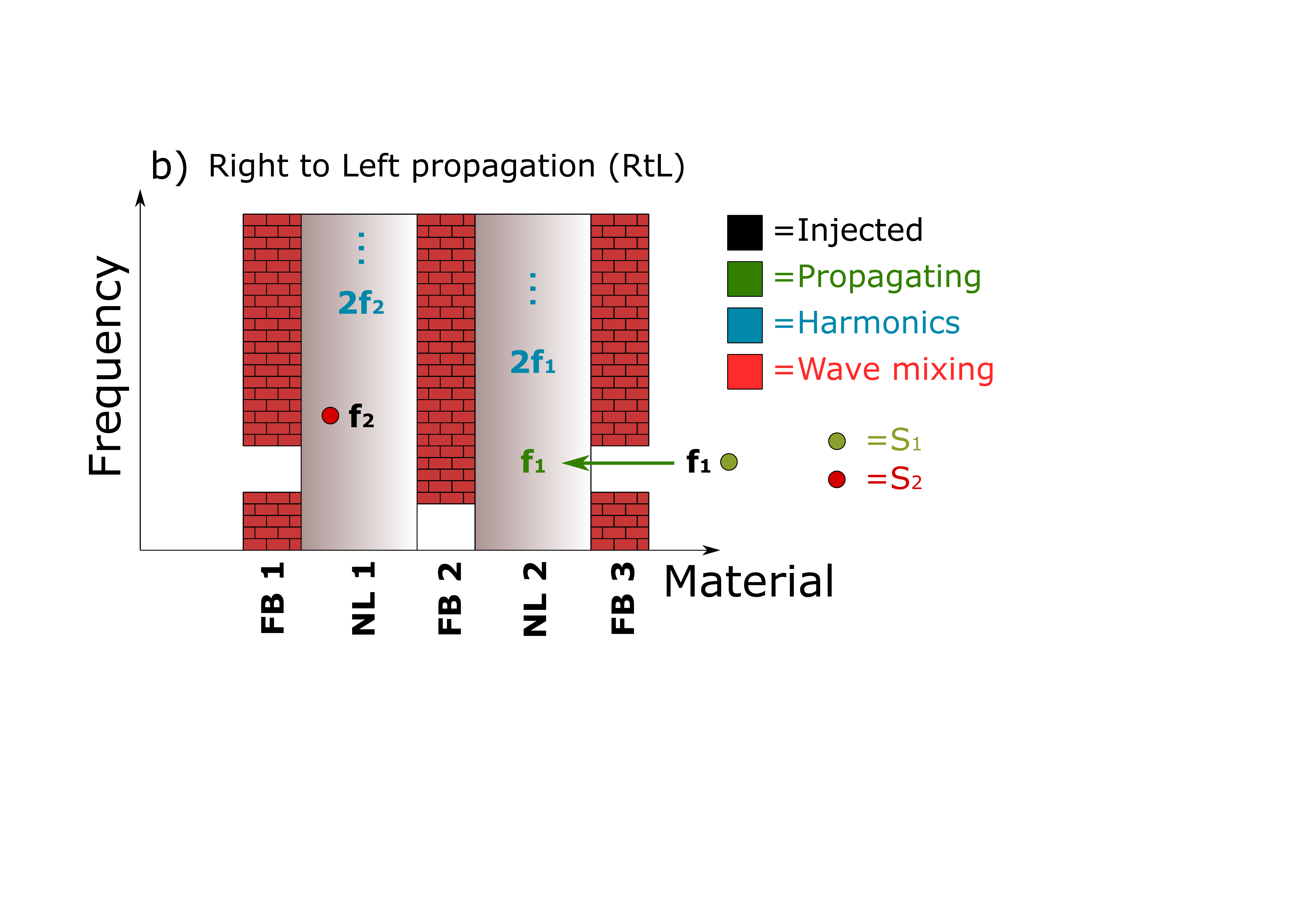}}
\subfigure
{\includegraphics[trim=0mm 40mm 0mm 10mm, clip=true, width=.98\textwidth]{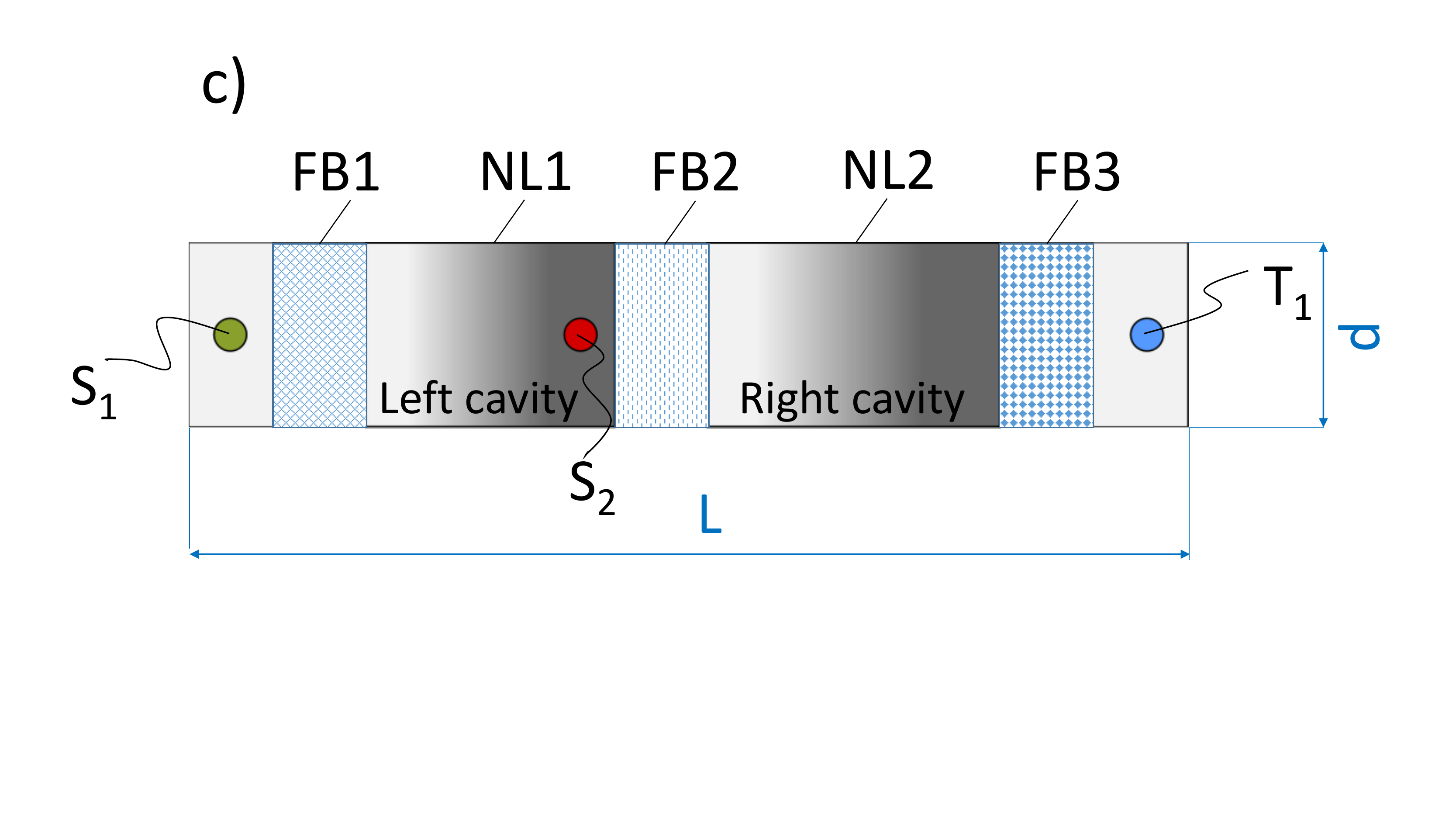}}
\end{minipage}
\caption{Schematic representation of the basic concept of the proposed AD for (a) left to right and (b) right to left propagation, respectively. $f_1$ and $f_2=\frac{3}{2}f_1$ are the injected wave components, while $f_0=\frac{1}{2}f_1$ is generated by wave mixing. The red barriers represent the frequency BGs, the gray zones are the nonlinear cavities (NL1 and NL2) where harmonic generation and wave mixing take place. The schematic representation of the numerical sample is represented in subplot (c).}\label{fig1}
\end{figure}

The working principle of the AD proposed in this study is illustrated in Fig. \ref{fig1} and can be described as follows:

\textit{i) Propagation from left to right (LtR, Fig. \ref{fig1}a)}: an input signal is injected (from $S_1$) into the device where it encounters a passband filter FB1 that selects a range of frequencies around $f_{1}$. These waves then travel through a first nonlinear elastic zone, named NL1, where a second frequency $f_2=\frac{3}{2}f_1$ can be injected from the source $S_2$. In this case, the presence of nonlinearity generates higher harmonics and the sum and difference frequencies (wave mixing), including $f_2-f_1=f_0=\frac{f_1}{2}$, which is a subharmonic of $f_1$. The next portion of the device, FB2, is a low-pass filter, which eliminates frequencies above $f_{0}$, and a second nonlinear zone, NL2, where the second harmonic $f_1=2f_0$ is generated. Finally, another passband filter (FB3) filters out \textit{f$_{0}$} and the harmonics higher than $f_1$, giving an overall output signal $f_{1}$  at the same frequency of the input.

\textit{ii) Propagation from right to left (RtL, Fig. \ref{fig1}b):} in this case, the input signal at $f_{1}$ travels through FB3 and through NL2 where higher harmonics are generated (but not $f_{0}$), and where no wave mixing process takes place (this breaks spatial reciprocity). The  next portion of the device, FB2, filters out the full signal, so that no signal propagates through NL1 and FB1, generating no output from the device. Notice that the source $S_2$ is present both in the forward and in the backward propagation (in this sense the AD is time invariant in its physical characteristics) and its role is to break spatial symmetry in the device. This mechanism allows us to overcome some of the difficulties in the practical realization encountered in other theoretical works that propose frequency-preserving ADs \cite{Chang2015,Zhong2016,Nassar2017}.  The present model/configuration has been conceived for monochromatic inputs, as usually done for nonlinearity-based ADs. More complicated designs can be considered by imposing a non-monochromatic wave injected by the source $S_2$. However, this is beyond the scope of this work. 

\section{Results}

To verify the feasibility and functionality of the device, we will first simulate its behaviour numerically (Figs.\ref{fig2}-\ref{fig4}) and then discuss its experimental realization (Figs.\ref{fig5} and \ref{fig6}). 

\subsection{Numerical verification}
In the numerical simulation (Fig.\ref{fig1}c), we model the device as an Aluminum plate with mass density ${\rho }_1=2700\,\mathrm{ kg/m^3}$, Young modulus $E=70\, \mathrm{GPa}$, and Poisson ratio $\nu=0.33$ and in-plane dimensions $L=105\,\mathrm{mm}$ and $d=6.6\,\mathrm{mm}$. The core of the device, in which reciprocity is broken, is composed by two nonlinear zones (NL1 and NL2 in Fig.\ref{fig1}), separated by a metamaterial (FB2).

The nonlinear sections NL1 and NL2 are realized by considering a zone of diffuse nonlinearity, and the numerical nonlinear parameters are set in order to produce about $10\%$ of harmonics and subharmonics. These two nonlinear zones are placed between two filters made of metamaterials or phononic crystals, which confine the frequency components of the wavefield falling in their BGs, creating a sort of resonant cavity (also named left and right cavities in the following). The dimensions of these regions and of the nonlinear elements can be tailored to enhance the desired frequencies through resonance effects ($f_0$ in the left, and $f_1$ in the right cavity). A nonclassical nonlinear model \cite{Ulrich2007,Gliozzi2014,Delsanto2003}, implemented using a Preisach-Mayergoyz \cite{Mayergoyz1985} space representation, is  adopted to simulate the nonlinear elastic response of these zones \cite{SupMat}.

The structure of each metamaterial/phononic crystal part (FB1-FB3) is described in detail in the Supplementary Material \cite{SupMat} together with its dispersion characteristics. The scalability of the results is guaranteed by the fact that the geometry of the constituent elements can easily be tuned to shift the pass bands to the desired frequencies.

\begin{figure}
\centering
\begin{minipage}[]{1\linewidth}
\subfigure
{\includegraphics[trim=0mm 30mm 10mm 40mm, clip=true, width=.98\textwidth]{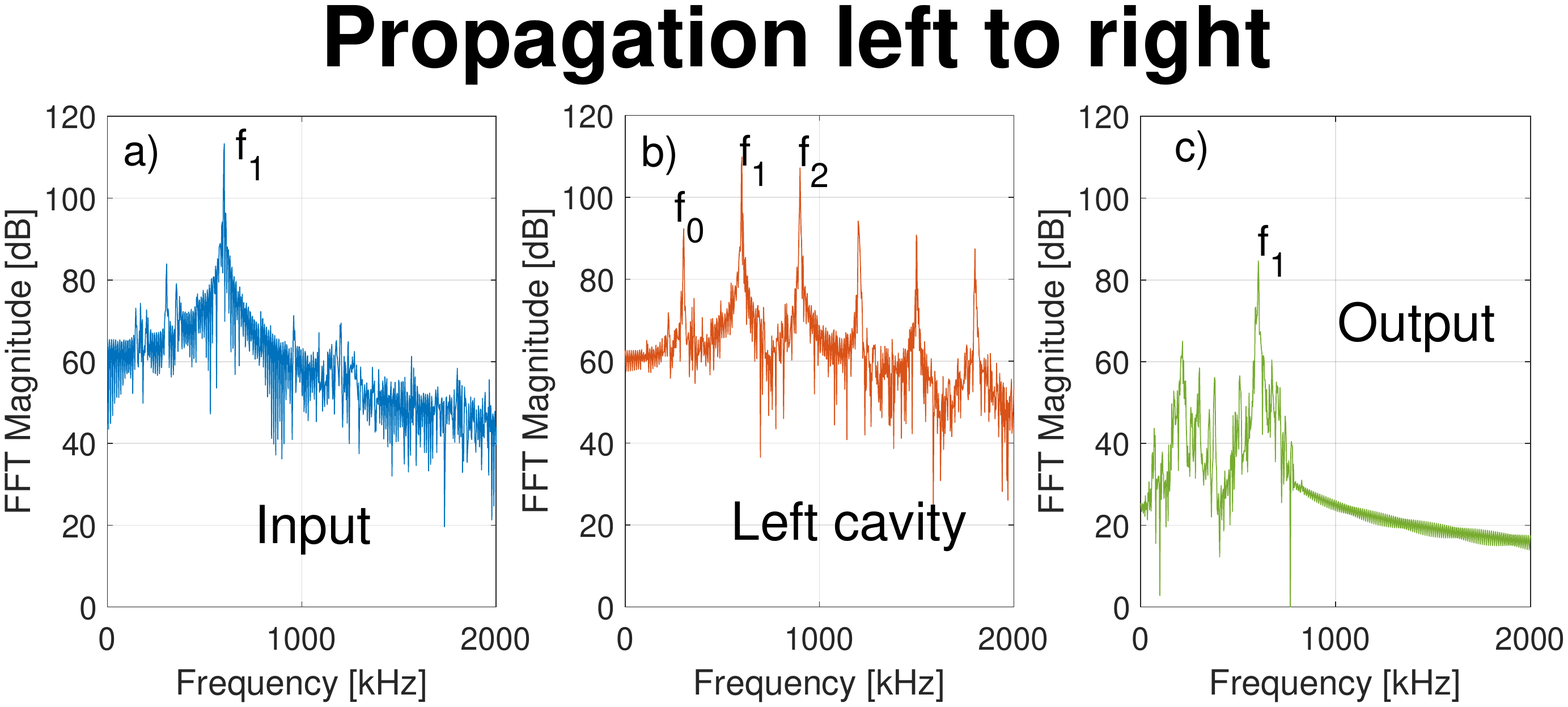}}
\subfigure
{\includegraphics[trim=0mm 30mm 10mm 30mm, clip=true, width=.98\textwidth]{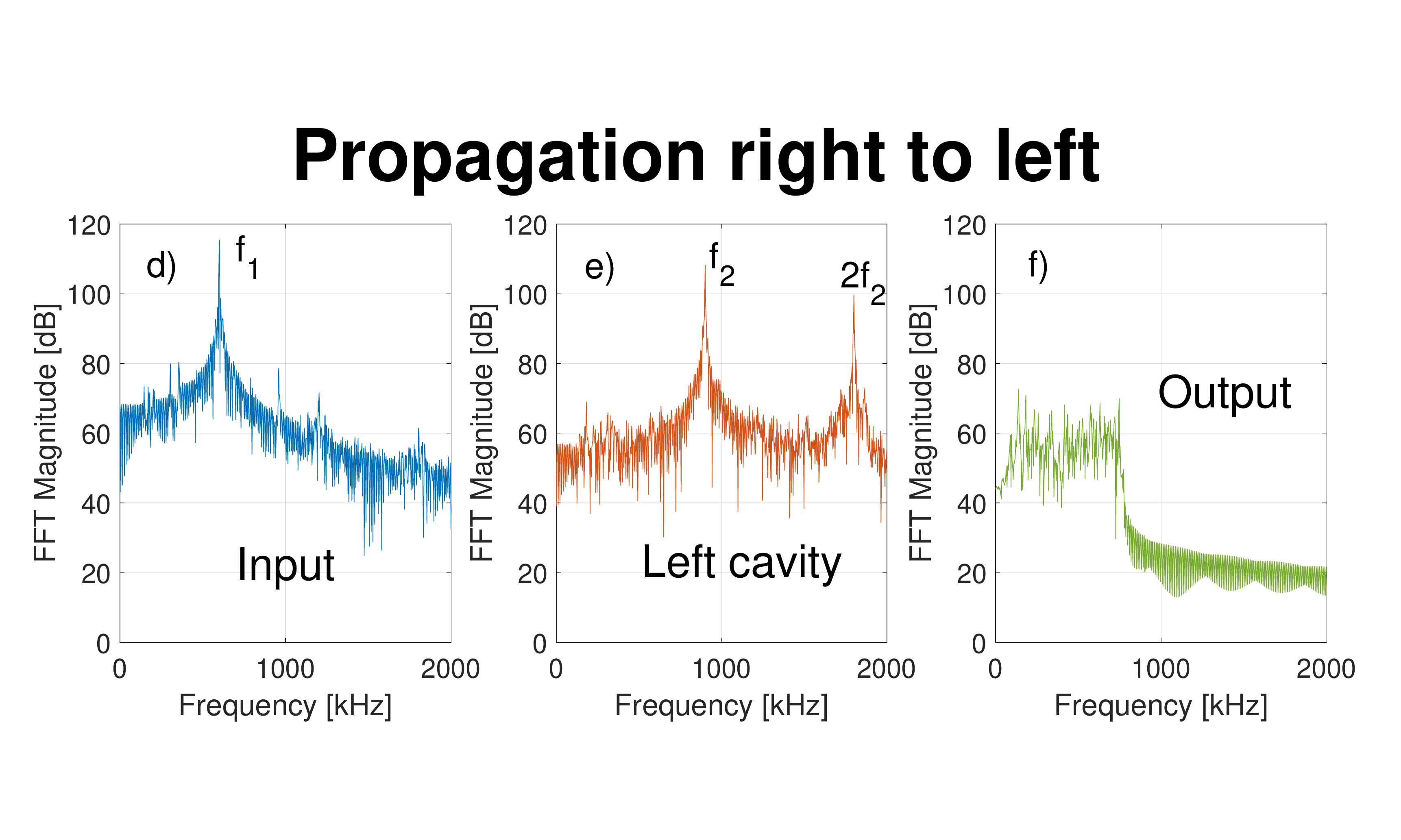}}
\end{minipage}
\caption{Fast Fourier Transforms of the signal recorded in the input (a, d), in the first cavity on the left (b, e) and at the output (c, f), for the two propagation directions (LtR and RtL in the first and second rows, respectively).}
\label{fig2}
\end{figure}

The excitation signal (a sinusoidal wave) is uniformly applied at the left boundary of FB1 (for LtR propagation) or at the right boundary of FB3 (for RtL propagation). We assume reflecting condition at the boundary that are free from excitation.

With this configuration we perform wave propagation simulations to demonstrate the effective feasibility of the AD. For the LtR (RtL) propagation, we inject a monochromatic wave of frequency $f_1=600\,\mathrm{kHz}$ on the left (right) side of the device and the corresponding $f_2=900\,\mathrm{kHz}$ in the left cavity. The output signal is recorded on the left (right) side of the sample (T1 in Fig-\ref{fig1}c). Fig. \ref{fig2} shows the Fast Fourier Transform (FFT) of the signals for LtR (a-c) and RtL (d-f) respectively. The signals are recorded at the input of the device (a,d), in the first cavity on the left (b,e) and at the output (c,f). While $f_1$ propagates from LtR, no signal is detected at the receiver when the propagation is in the other direction. The difference between the two cases (reported in the upper and lower parts of Fig. \ref{fig2}, respectively) lies in the generation in the left cavity of the frequency $f_0$, which is the only component that can propagate from NL1 to NL2. 

Although any mechanism able to generate sub-harmonics \cite{alippi1992, Bosia2006} of $f_1$ can be appropriate, the mechanism based on wave mixing adopted here to generate $f_0$ has several
advantages. The first is that wave mixing is an extremely efficient way to produce sub-harmonics and no threshold mechanism seems to be at play. Moreover, the source $S_2$ can be tuned in order to decrease or increase the amplitude of the $f_0$ component, and in the limit case to suppress it. Thus, the device can be used as an \textit{on-off} or an \textit{amplitude-tuning} switch.
Two different simulations are presented to demonstrate these applications. 

\begin{figure}
\centering
\includegraphics[trim=0mm 00mm 0mm 00mm, clip=true, width=.50\textwidth]{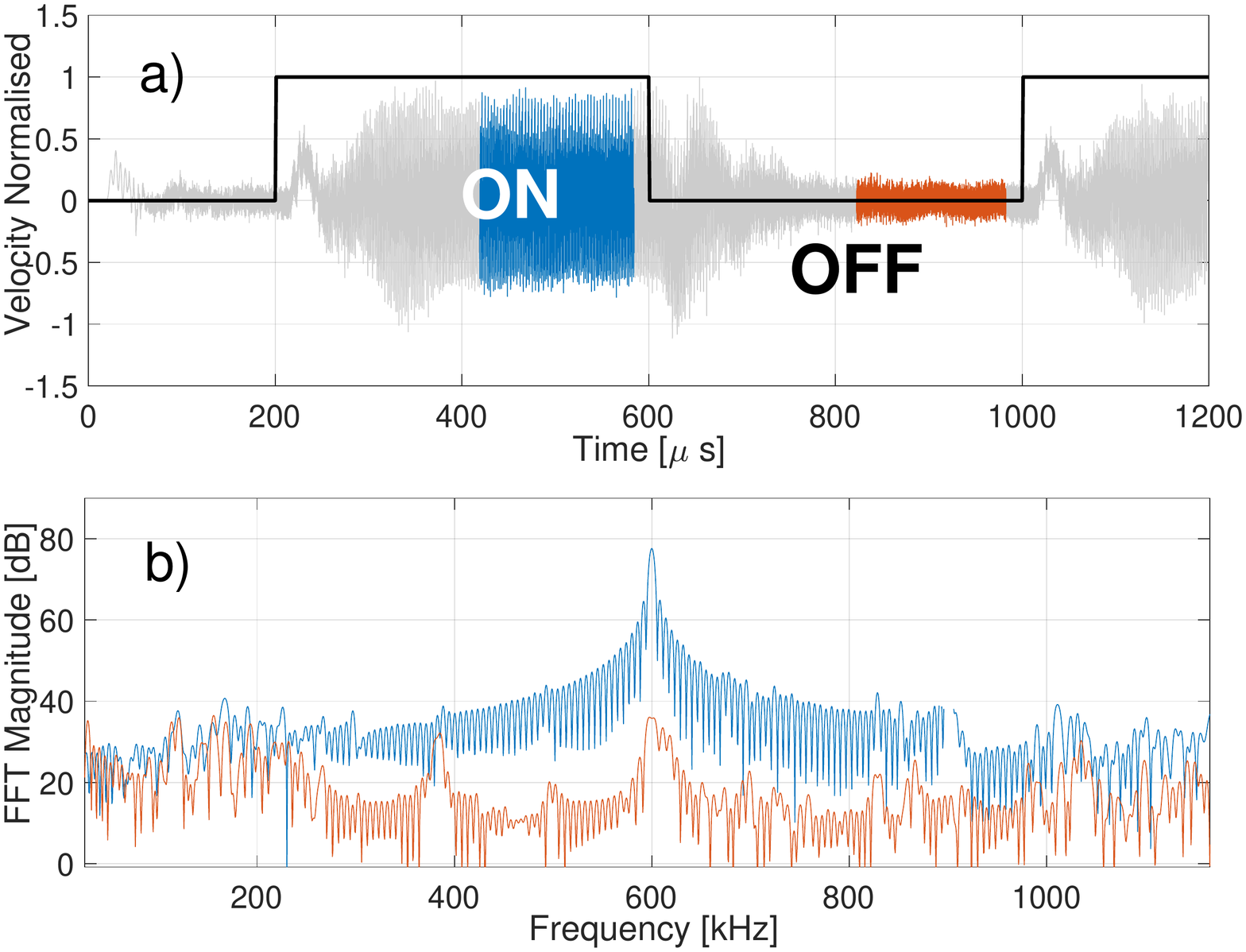}
\caption{Acoustic on-off switch. By switching on/off the source $S_2$, the wave generated by $S_1$ can/cannot propagate through the device. This is visible both in the a) time and b) frequency domain. The FFT performed over different time windows (highlighted with different colors in subplot (a)) shows the different frequency content of the propagating wave.}
\label{fig3}
\end{figure}

In the first case, the source $S_2$ (the pump) is switched on/off at regular time intervals  and the corresponding output recorded (see Fig.\ref{fig3}). As shown in Fig. \ref{fig3}a, the signal is prevented from propagating when the source $S_2$ is switched off. This is also evident in the FFT analysis performed by windowing the time signal for the two different cases ($S_2$ on/off in Fig. \ref{fig3}b). This demonstrates the use of the AD as a switch.

\begin{figure}
\centering
{\includegraphics[trim=10mm 25mm 20mm 30mm, clip=true, width=.48\textwidth]{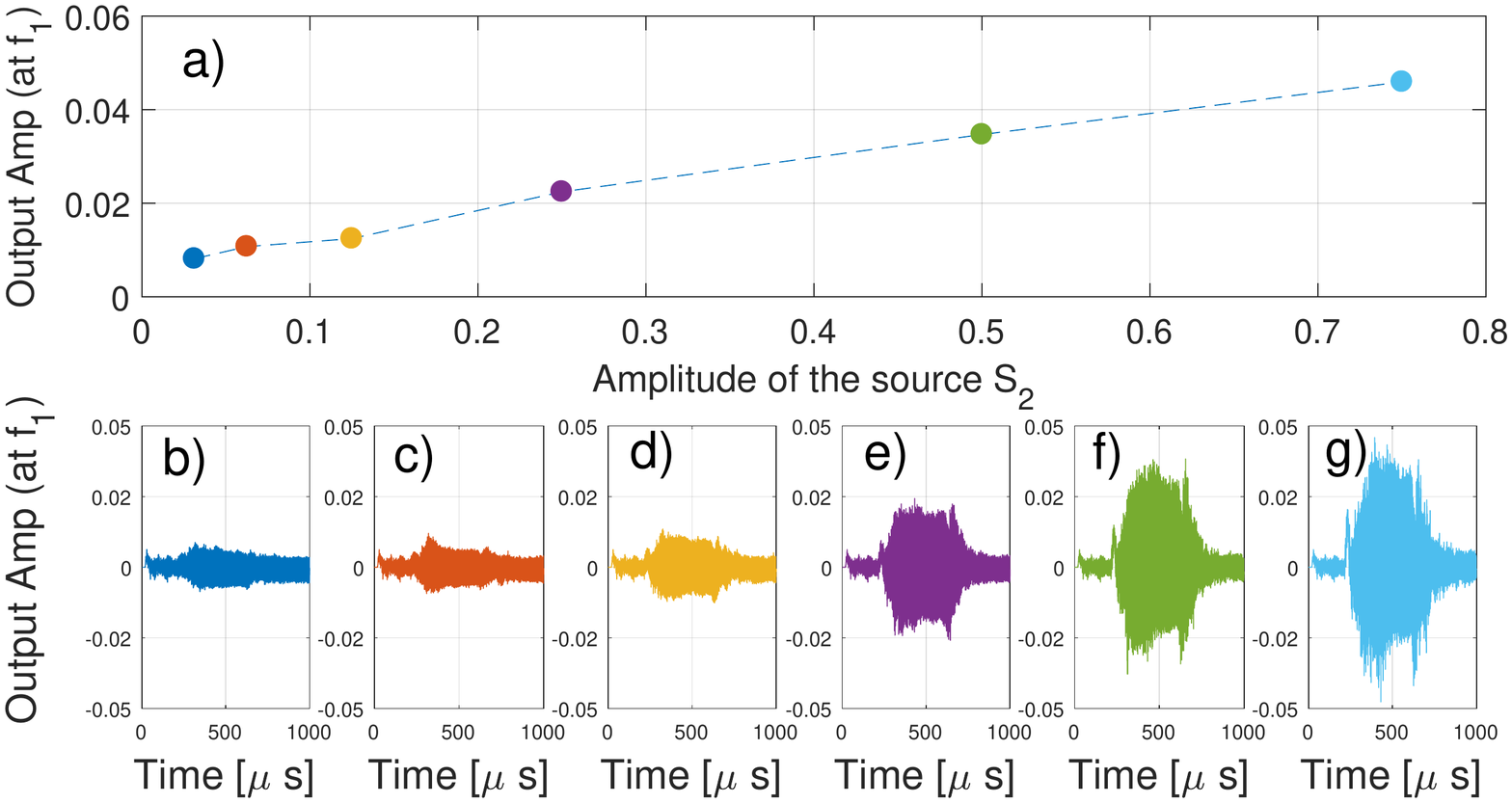}}
\caption{Amplitude-tuning  switch. (a) Increase in output amplitude as a function of the pump amplitude $S_2$, for constant input amplitude ($S_1$); (b-g) Corresponding outputs signals.}
\label{fig4}
\end{figure}

The same numerical experiment is then repeated at increasing amplitude of the pump ($S_2$) while keeping the amplitude of $S_1$ fixed. Since the amplitude of the $f_0$ component (the subharmonic of the input) is proportional to the product of the two mixed frequency amplitudes, it is possible to change  the output signal amplitude by tuning the amplitude of the pump ($S_2$, in this case), as shown in Fig. \ref{fig4}. 
This generates the possibility to realize a switch with a variable amplitude output. Moreover, from a theoretical point of view this opens the way to the possibility of considerably increasing the efficiency of the device by pumping energy from $S_2$ and increasing the output amplitude at will. From a practical point of view, a large amplification may be limited by spurious nonlinear effects and by the large amount of energy required. This limitation could be partially overcome by finding a very efficient nonlinear system. Work is in progress on this aspect.

\subsection{Experimental realization}

\begin{figure}
\centering
{\includegraphics[trim=0mm 30mm 0mm 0mm, clip=true, width=.48\textwidth]{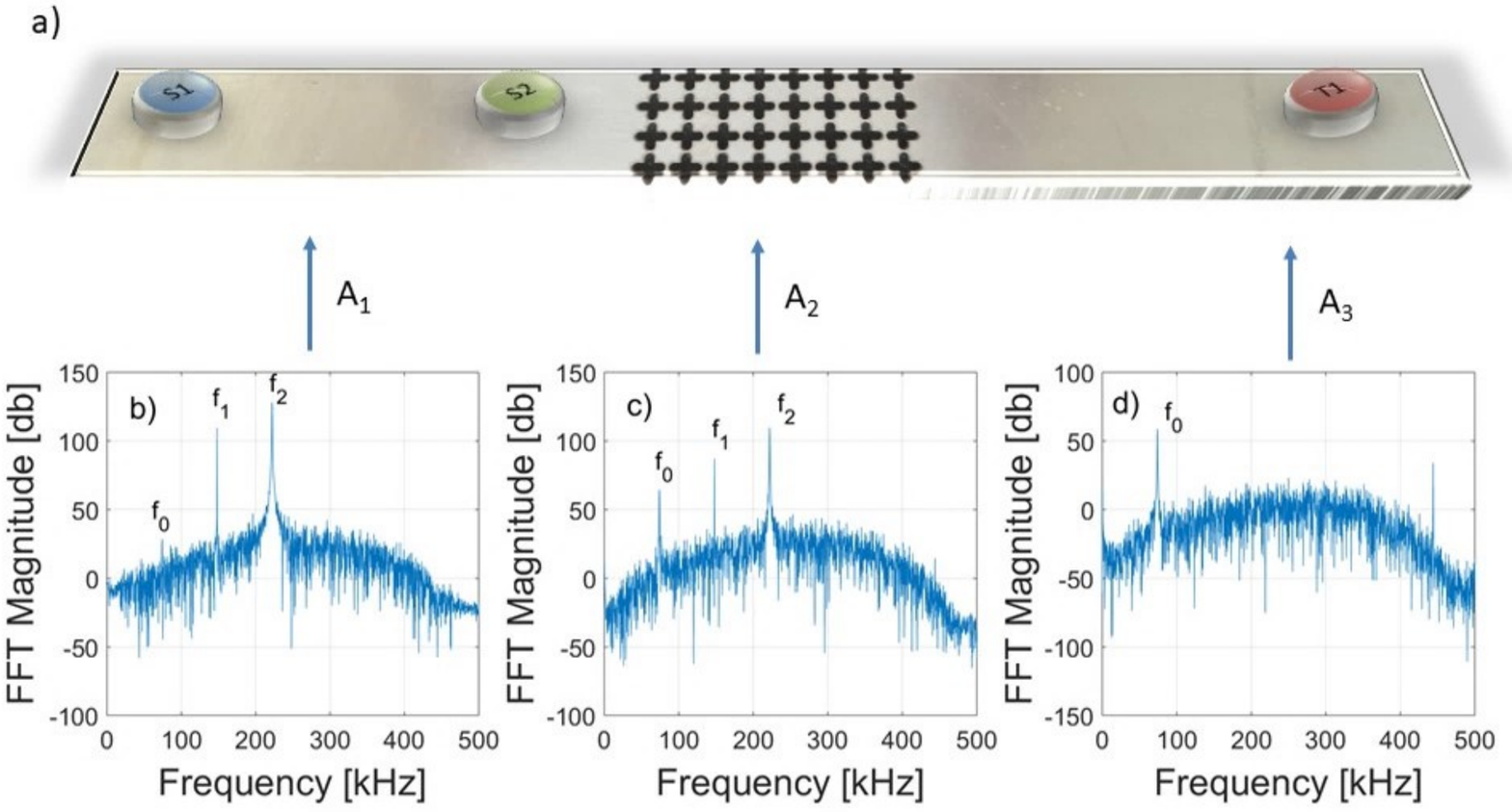}}
\caption{Experimental realization of AD: a) central part of the device with the two nonlinear cavities and the central barrier. The positions of the two transducers $S_1$ and $S_2$ and of the receiver $T1$ are reported. (b-d) Spectral content detected with a laser scan in three different points ($A_1$, $A_2$ and $A_3$) of the device in each of the three zones indicated.}
\label{fig5}
\end{figure}

The discussed design of the AD is quite general and can be realized with different nonlinearity types, filtering characteristics or optimized properties. We demonstrate  its feasibility through the experimental realization of a prototype, representing the central part of the device, which is responsible for the breaking of reciprocity.
We use a $380 \times 40 \times 6$ mm$^3$ aluminium plate ($\rho = 2700$ kg$/$m$^3$, $E = 70$ GPa and $\nu = 0.33$) with a phononic crystal region representing the filtering barrier (FB2 in Fig.\ref{fig1}). 
The phononic crystal is located between two regions that represent the left (with NL1) and right (with NL2) cavities in Fig. \ref{fig1}. 
FB2 consists of a 2D array of $4 \times 8$ cross-like cavities \cite{Miniaci2015}, fabricated using waterjet cutting, with a lattice parameter of $a = 10$ mm (see \cite{SupMat} for geometrical details).  Dimensions have been designed so to suppress frequencies from $124$ kHz to $175$ kHz and $191$ kHz to $236$ kHz, in the propagation from one cavity to the other (see \cite{SupMat} for further details).
It follows that the working frequency of this AD is $f_1=150\,\mathrm{kHz}$, while the pump $S_2$ needs to be set at a frequency $f_2=225\,\mathrm{kHz}$. In this simplified realization, in the LtR propagation, the two sources ($S_1$ and $S_2$) are located in the same cavity on the left, while the receiver ($T_1$)is situated in the right cavity, as shown in Fig. \ref{fig5}a. 
The nonlinearity is generated in the two cavities, by superposing onto the plate a small object coupled with a drop of water \cite{Miniaci2017}. The clapping of the surfaces, due to the action of the elastic wavefield propagating in the plate gives rise to typical nonlinear effects (i.e. harmonics and wavemixing). 

In the experiments, the emitting piezoelectric contact transducer was connected to an arbitrary waveform generator (Agilent 33500B) through a $50$dB linear amplifier (FLC Electronics A400). The receiving transducer/laser interferometer was connected to an oscilloscope (Agilent Infiniium DSO9024H) for data acquisition.

Figs. \ref{fig5}(b-d) show the change in the spectral content of the signal at three different points in the sample detected by the laser vibrometer: only the nonlinearity in the first cavity  (NL1) is activated (no reciprocity breaking is expected in this case), so the only frequency that can travel from left to right is the frequency $f_0=75\,\mathrm{kHz}$, generated by wave mixing in the cavity on the left (see \cite{SupMat}). 

\begin{figure}
\centering
\begin{minipage}[]{1\linewidth}
\subfigure
{\includegraphics[trim=10mm 55mm 10mm 50mm, clip=true, width=1.\textwidth]{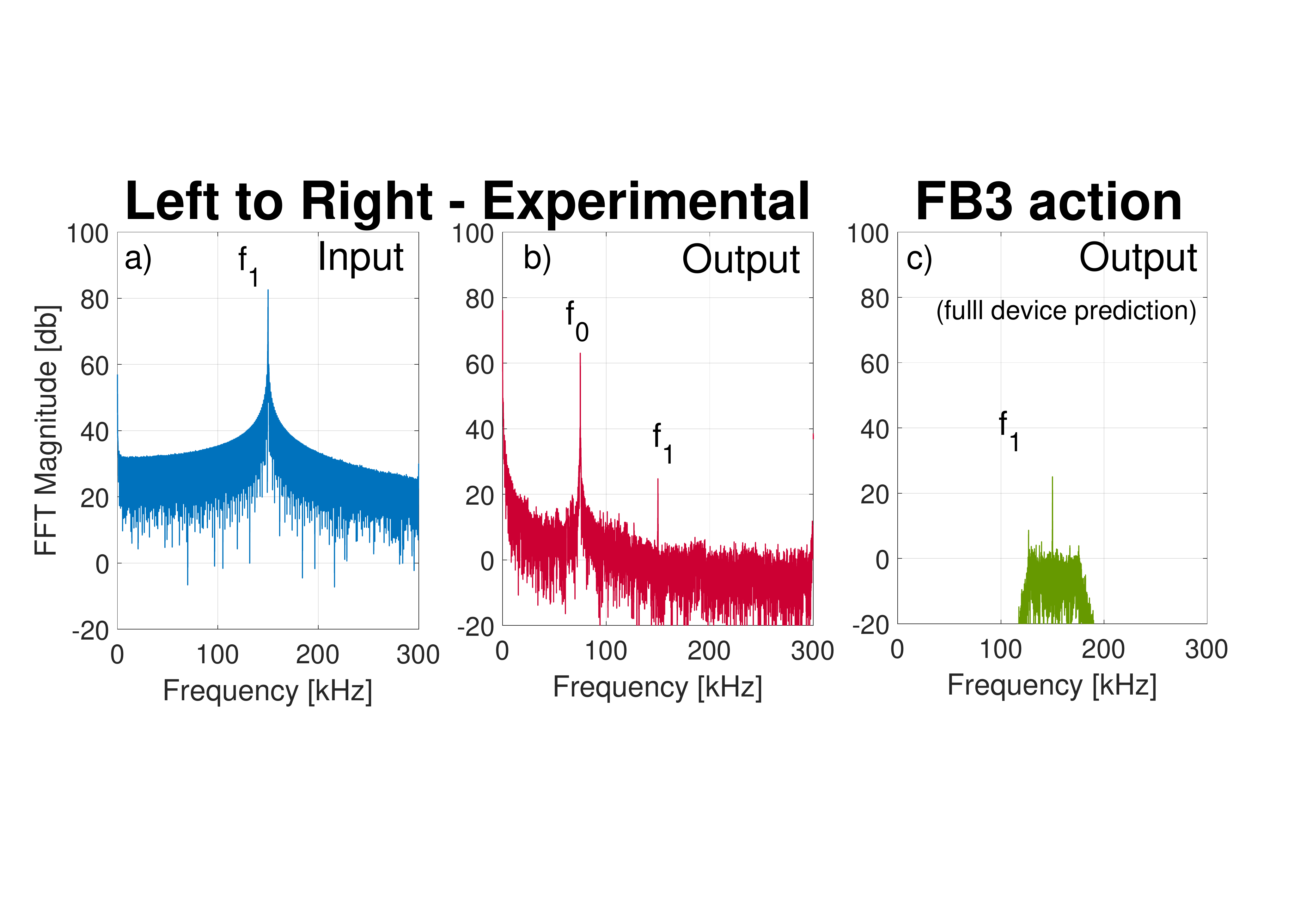}}
\subfigure
{\includegraphics[trim=10mm 55mm 10mm 50mm, clip=true, width=1.\textwidth]{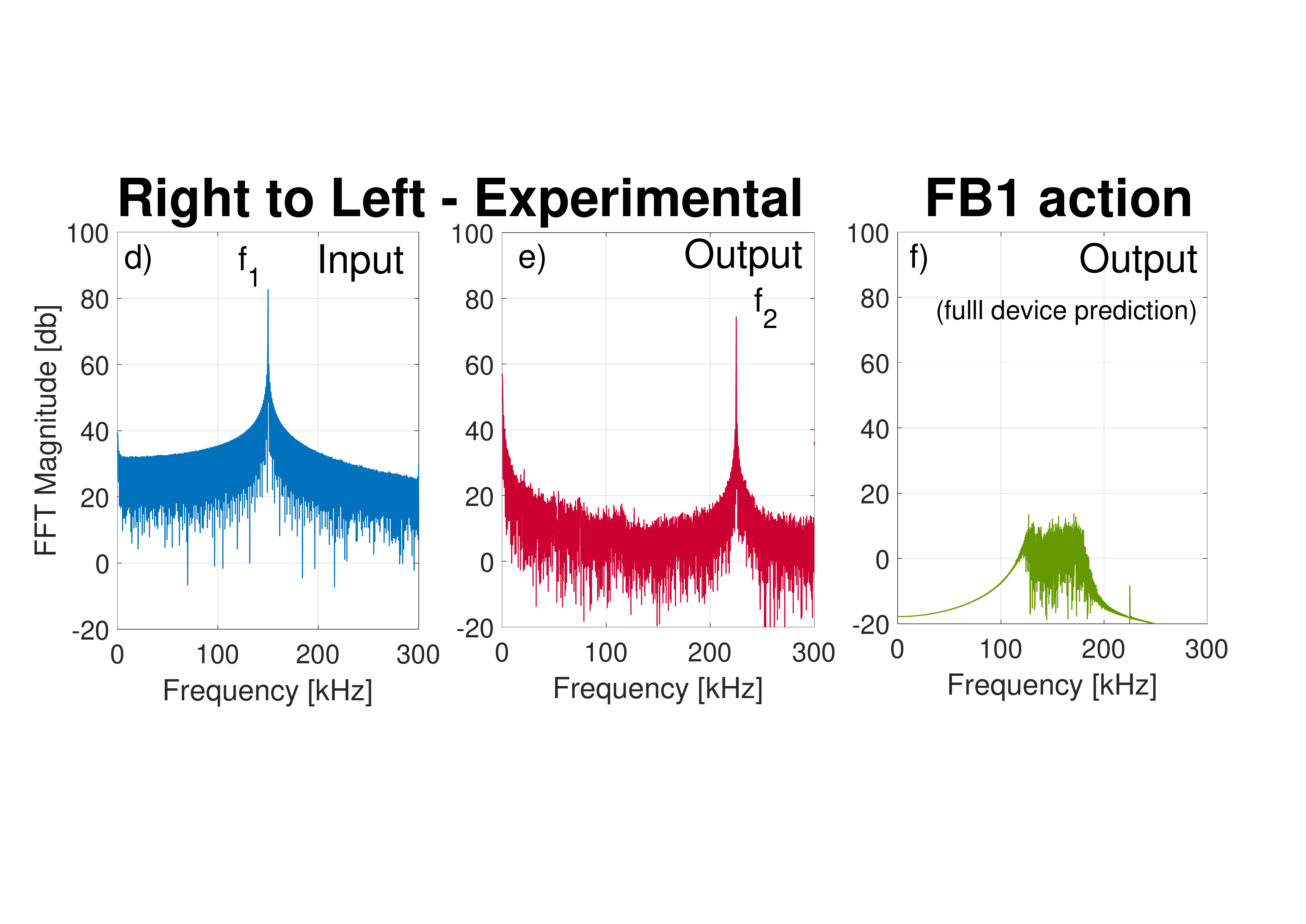}}
\end{minipage}
\caption{Experimental results for LtR (a,b) and RtL (d,e) propagation. In the first and second columns the FFT of the injected signal (a,d) and of the output (b,e) are reported, respectively. The third column represents the simulated effect on the FFT of the output of the phononic barriers in the full device.}
\label{fig6}
\end{figure}

In the second experiment (Fig. \ref{fig6}), the nonlinearity is activated in both cavities in order to demonstrate breaking of reciprocity, and wave propagation is studied in both directions of the device by placing the transducer $S_1$ in one of the two cavities and the receiver $T_1$ in the other, and subsequently inverting them, leaving the position of $S_2$ unchanged. The filtering action of FB1 and FB3 in the complete device is realized here in post-processing by imposing a numerical band-pass filter (centered around $f_1=150\,\mathrm{kHz}$) to the output signals (third column in Fig. \ref{fig6}). Despite the relatively small amplitudes, the symmetry breaking in the wave propagation for the frequency $f_1$ is evident in the two cases reported in the two rows of Fig. \ref{fig6}. The FFT of the input signal injected at the source $S_1$ and of the output are shown in the first and in the second columns, respectively. In the third column the output is band-pass filtered, simulating the action of the phononic barriers. The difference in the output obtained in the left and right propagation demonstrates the functionality of the AD.

\section{Conclusions}
In summary, we have presented numerical and experimental results demonstrating the feasibility of an acoustic diode based on alternating nonlinear elastic and metamaterial frequency-filtering regions, with time-invariant, frequency preserving characteristics. The design concept is sufficiently general to allow flexibility in its realization, involving different combinations of nonlinearity and BG mechanisms, and the use of phononic crystals or resonant metamaterials provides the opportunity to tune and scale results to the desired device sizes and frequency ranges. Additionally, the adoption of an input monochromatic driving signal allows the adaptation of the concept to different types of devices, such as switches or transistors, which can be exploited in practical applications in the field of acoustics or ultrasonics \cite{Bilal2017}. These can potentially be coupled and integrated with recently introduced metamaterial-based sensors for damage detection and localization \cite{Miniaci2017} or for other advanced signal manipulation purposes. However, for an integrated use in a more advanced apparatus, an improvement in the efficiency and stability of the experimental results is required. For this, improved nonlinear elastic solutions are currently under study.

\section{Acknowledgments}

\noindent M. M. has received funding from the European Union’s Horizon 2020 research and innovation program under the Marie Skłodowska-Curie Grant Agreement No. 754364. A. K. acknowledges financial support from the Department of Department of Civil, Environmental and Mechanical Engineering, University of Trento. NMP is supported by the European Commission under the Graphene Flagship Core 2 grant No. 785219 (WP14 “Composites”) and FET Proactive ``Neurofibres'' grant No. 732344 as well as by the Italian Ministry of Education, University and Research (MIUR) under the ``Departments of Excellence'' grant L.232/2016. FB is supported by ``Neurofibres'' grant No. 732344  and by project ``Metapp'', (n. CSTO160004) cofunded by Fondazione San Paolo. 



\begin{thebibliography}{47}%
\makeatletter

\bibitem{Fink2000} M. Fink, D. Cassereau, A. Derode, C. Prada, P. Roux, M. Tanter, J.-L. Thomas and F. Wu, Rep. Progr. Phys., {\bf 63}, 1933--1995 (2000).

\bibitem{Liang2009} B. Liang, B. Yuan and J.-C. Cheng, Phys. Rev. Lett., {\bf 103}, 104301 (2009). 

\bibitem{Liang2010} B. Liang, X.-Y. Zou, D. Zhang and J.-C. Cheng, Nat. Mater., {\bf 9}, 989--992 (2010). 

\bibitem{Li2010} B. Li, Nat. Mater., {\bf 9}, 962--963 (2010).

\bibitem{Haar2006} G. ter Haar, C. Coussios, Int. J. Hyperthermia, {\bf 23}(2), 89--104 (2006)

\bibitem{Liu2016} K. Liu, S. He, Scien. Rep., {\bf 6}, 30206 (2016)

\bibitem{Maznev2013} A. Maznev, A. Every and O. Wright, Wave Motion, {\bf 50}, 776--784 (2013).  

\bibitem{Trainiti2016} G. Trainiti and M. Ruzzene, New J. Phys., {\bf 18}, 083047 (2016).

\bibitem{Popa2014} B. I. Popa and S.A. Cummer, Nat. Commun. {\bf 5}, 3398 (2014).

\bibitem{Lepri2011} S. Lepri and G. Casati, Phys. Rev. Lett., {\bf 106}, 164101 (2011). 

\bibitem{Li2011} X.-F. Li, X. Ni, L. Feng, M.-H. Lu, C. He and Y.-F. Chen, Phys. Rev. Lett., {\bf 106}, 084301 (2011). 

\bibitem{Feng2014} F.Li, P.Anzel, J. Yang, P.G. Kevrekidis and C. Daraio, Nat. Comm.,{\bf 5}, 5311 (2014).

\bibitem{Fleury2014} R. Fleury, D. L. Sounas, C. F. Sieck, M. R. Haberman and A. Al\`{u}, Science, {\bf 343}, 516--519 (2014). 

\bibitem{Sun2012} H.-X. Sun, S.-Y. Zhang and X.-J. Shui, Appl. Phys. Lett., {\bf 100}, 103507 (2012). 

\bibitem{Zhu2016} Y.-F. Zhu, Z.-M. Gu, B. Liang, J. Yang, J. Yang, L.-L. Yin and J.-C. Cheng, Appl. Phys. Lett., {\bf 109}, 103504 (2016). 

\bibitem{Sklan2015} S. R. Sklan, AIP Advances, {\bf 5}, 053302 (2015). 

\bibitem{Ma2013} C. Ma, R. G. Parker and B. B. Yellen,  J. Sound and Vibration, {\bf 332}, 4876--4894 (2013). 

\bibitem{Morvan2010} B. Morvan, A. Tinel, A.-C. Hladky-Hennion, J. Vasseur and B. Dubus, Appl. Phys. Lett., {\bf 96}, 101905 (2010). 

\bibitem{Kushwaha1993} M. S. Kushwaha, P. Halevi, L. Dobrzynski and B. Djafari-Rouhani, Phys. Rev. Lett., {\bf 71}, 2022 (1993). 

\bibitem{Martinez1995} R. Martinez-Sala, J. Sancho, J. V. Sanchez, V. Gomez, J. Llinares and F. Meseguer,  Nature, {\bf 378}, 241 (1995). 

\bibitem{Fraternali2017} F. Fraternali, G. Carpentieri and A. Amendola, J. Mech. Phys. Solids, {\bf 99}, 259-271, (2017). 

\bibitem{Yang2004} S. Yang, J. H. Page, Z. Liu, M. L. Cowan, C. T. Chan and P. Sheng, Phys. Rev. Lett., {\bf 93}, 024301 (2004). 

\bibitem{Brun2010} M. Brun, S. Guenneau, A. B. Movchan and D. Bigoni, J. Mech. Phys. Solids, {\bf 58}, 1212 (2010). 

\bibitem{Gliozzi2015} A. S. Gliozzi, M. Miniaci, F. Bosia, N. M. Pugno and M.Scalerandi,  Appl. Phys. Lett., {\bf 107}, 161902 (2015). 

\bibitem{Miniaci2017} M. Miniaci, A. S. Gliozzi, B. Morvan, A. Krushynska, F. Bosia, M. Scalerandi, and N. M. Pugno, Phys. Rev. Lett. {\bf 118}, 214301 (2017).

\bibitem{Miniaci2018}  M. Miniaci, R. K. Pal, B. Morvan, and M. Ruzzene, Phys. Rev. X, Phys. Rev. X {\bf 8}, 031074 (2018).

\bibitem{Tsakmakidis2017} K.L.Tsakmakidis, L. Shen, S. A. Schulz, X. Zheng, et al., Science, {\bf 356}, 1260 (2017)

\bibitem{Zhang2011}  S. Zhang, C. Xia, and N. Fang, Phys. Rev. Lett., {\bf 106}, 024301 (2011). 

\bibitem{Pasini2015}  N. Nadkarni, C. Daraio, and D.M. Kochmann, Phys. Rev. E 90, 023204 (2014).

\bibitem{Bertoldi2017} Bertoldi K. , Annu. Rev. Mater. Res., {\bf 47}, 51-–61 (2017).

\bibitem{Wang2016} Wang, Y.Z., Li, F.M., Wang, Y.S., Int. J. Mech. Sci., {\bf 106}, 357--362 (2016).

\bibitem{Scalerandi2012} M. Scalerandi, A. S. Gliozzi and C. L. E. Bruno, J. Acoust. Soc. Am., {\bf 131},  EL81 (2012). 

\bibitem{Deymier2013} P. Deymier, Acoustic Metamaterials and Phononic Crystals, Berlin: Springer, 2013. 

\bibitem{Krushynska2017} A. Krushynska, M. Miniaci, F. Bosia and N. Pugno, Ext. Mech. Lett., {\bf 12}, 30--36 (2017). 

\bibitem{Carrara2013} M. Carrara, M. Cacan, J. Toussaint, M. Leamy, M. Ruzzene and A. Erturk, Smart Mater. Struct., {\bf 22}, 065004 (2013).

\bibitem{Chang2015} C. Liu, Z. Du, Z. Sun, H. Gao, and X. Guo, Phys. Rev. Appl. {\bf 3}, 064014 (2015).

\bibitem{Zhong2016} Z.-m. Gu, J. Hu, B. Liang, X.-y. Zou, J-c. Cheng, Sci. Rep. {\bf 6}, 19824 (2016).

\bibitem{Nassar2017} H. Chen, A.N. Norris, M.R. Haberman, G.L. Huang, Proc. R. Soc. A {\bf 473}, 20170188 (2017).

\bibitem{Ulrich2007} T. J. Ulrich, P. A. Johnson, R. A. Guyer, Phys. Rev. Lett., {\bf 98}, 104301 (2007).

\bibitem{Gliozzi2014} A. S. Gliozzi and M. Scalerandi, J. Acoust. Soc. Am., {\bf 136} (4), 1530 (2014).

\bibitem{Delsanto2003} P. P. Delsanto, M. Scalerandi, Phys. Rev. B, {\bf 68}, 064107 (2003).

\bibitem{Mayergoyz1985} I. D. Mayergoyz, J. Appl. Phys., {\bf 57}, 3803 (1985).

\bibitem{SupMat} See supplemental material for further details.

\bibitem{alippi1992} A. Alippi, G. Shkerdin, A. Bettucci, F. Craciun, E. Molinari, and A. Petri, Phys. Rev. Lett., {\bf 69}, 3318 (1992) 

\bibitem{Bosia2006} F. Bosia, N. Pugno, A. Carpinteri, Wave Motion {\bf 43}, 689–-699 (2006). 








\bibitem{Bilal2017} O. R. Bilal, A. Foehr, and C. Daraio, PNAS, {\bf 114}, 4603--4606 (2017).

\bibitem{Miniaci2015} M. Miniaci, A. Marzani, N. Testoni, L. De Marchi, Ultrasonics {\bf 56}, 251--259 (2015).

\end{thebibliography}
\end{document}